\documentclass[aps,prl,twoside,twocolumn,superscriptaddress,floatfix,nofootinbib,showpacs]{revtex4-1}
\usepackage{hyperref}
\usepackage{amsmath}
\usepackage{bm}
\usepackage{xcolor}
\usepackage{float}

 \usepackage{multirow}
  \usepackage{mathtools}

\newcommand\beq{\begin{equation}}
\newcommand\eeq{\end{equation}}
\newcommand\beqn{\begin{eqnarray}}
\newcommand\eeqn{\end{eqnarray}}

\newcommand{\ba}{\begin{eqnarray}}
\newcommand{\ea}{\end{eqnarray}}
\newcommand{\be}{\begin{equation}}
\newcommand{\ee}{\end{equation}}
\newcommand{\bal}{\begin{aligned}}
\newcommand{\eal}{\end{aligned}}

\newcommand\lsim{\mathrel{\rlap{\lower4pt\hbox{\hskip1pt$\sim$}}
        \raise1pt\hbox{$<$}}}
\newcommand\gsim{\mathrel{\rlap{\lower4pt\hbox{\hskip1pt$\sim$}}
        \raise1pt\hbox{$>$}}}

\newcommand{\jcap}{{J.~Cosm.~Astrop.~Phys.}}

\newcommand{\aap}{{Astron.~Astrophys.}}
\newcommand{\apjl}{{Astrophys.~J.~Lett.}}
\newcommand{\apjs}{{Astrophys.~J.~Supp.}}

\newcommand{\mnras}{{Mon.~Not.~R.~Astron.~Soc.}}
\usepackage{graphicx}
\usepackage[large]{subfigure}
\usepackage{amssymb, amsmath}
\usepackage[amssymb]{SIunits}
\usepackage{epstopdf}
\usepackage{aas_macros}
\usepackage{natbib}

\begin{document}

\title{Evidence for the kinematic Sunyaev-Zel'dovich effect with ACTPol and velocity reconstruction from BOSS}

%%%%%%%%%%%%%%%%%%%%%%%%%%%%%%%%%%%%%%%%%%%%%%%%%%%%%%%%%%%%%%%%%%%%%%%%%%%
\author{Emmanuel Schaan}\email{eschaan@astro.princeton.edu}
\affiliation{Dept.~of Astrophysical Sciences, Peyton Hall, Princeton University, Princeton, NJ USA 08544}
\author{Simone~Ferraro}
\affiliation{Dept.~of Astrophysical Sciences, Peyton Hall, Princeton University, Princeton, NJ USA 08544}
\affiliation{Miller Institute for Basic Research in Science, University of California, Berkeley, CA, 94720, USA}
\author{Mariana~Vargas-Maga\~na}
\affiliation{Instituto de Fisica, Universidad Nacional Aut\'onoma de M\'exico, Apdo. Postal 20-364, M\'exico}
\author{Kendrick~M.~Smith}
\affiliation{Perimeter Institute for Theoretical Physics, Waterloo, ON N2L 2Y5, Canada}
\author{Shirley~Ho}
\affiliation{Department of Physics, Carnegie Mellon University, 5000 Forbes Avenue, Pittsburgh, PA 15213, USA} 
\author{Simone Aiola}
\affiliation{Department of Physics and Astronomy, University of Pittsburgh, Pittsburgh, PA 15260 USA and
Pittsburgh Particle Physics, Astrophysics, and Cosmology Center, University of Pittsburgh, Pittsburgh PA 15260}
\author{Nicholas Battaglia}
\affiliation{Dept.~of Astrophysical Sciences, Peyton Hall, Princeton University, Princeton, NJ USA 08544}
\author{J.~Richard~Bond}
\affiliation{Canadian Institute for Theoretical Astrophysics, University of
Toronto, Toronto, ON, Canada M5S 3H8}
\author {Francesco De Bernardis}
\affiliation{Department of Physics, Cornell University, Ithaca, NY 14853, USA}
\author{Erminia~Calabrese}
\affiliation{Dept.~of Astrophysical Sciences, Peyton Hall, Princeton University, Princeton, NJ USA 08544}
\affiliation{Sub-Department of Astrophysics, University of Oxford, Keble Road, Oxford, UK OX1 3RH}
\author{Hsiao-Mei Cho }
\affiliation{SLAC National Accelerator Laboratory, 2575 Sandhill Hill Road, Menlo Park, CA 94025}
\author{Mark~J.~Devlin}
\affiliation{Department of Physics and Astronomy, University of
Pennsylvania, 209 South 33rd Street, Philadelphia, PA, USA 19104}
\author{Joanna~Dunkley}
\affiliation{Sub-Department of Astrophysics, University of Oxford, Keble Road, Oxford, UK OX1 3RH}
\author{Patricio~A.~Gallardo}
\affiliation{Department of Physics, Cornell University, Ithaca, NY 14853, USA}
\author{Matthew~Hasselfield}
\affiliation{Dept.~of Astrophysical Sciences, Peyton Hall, Princeton University, Princeton, NJ USA 08544}
\author{Shawn~Henderson}
\affiliation{Department of Physics, Cornell University, Ithaca, NY 14853, USA}
\author{J.~Colin~Hill}
\affiliation{Dept. of Astronomy, Pupin Hall, Columbia University, New York, NY, USA 10027}
\author{Adam D. Hincks}
\affiliation{UBC (University of British Columbia, Department of Physics and Astronomy, 6224 Agricultural Road, Vancouver BC V6T 1Z1, Canada)}
\author{Ren\'ee~Hlozek}
\affiliation{Dept.~of Astrophysical Sciences, Peyton Hall, Princeton University, Princeton, NJ USA 08544}
\author{Johannes Hubmayr}
\affiliation{National Institute of Standards and Technology, Boulder, CO USA 80305}
\author{John P. Hughes}
\affiliation{Department of Physics and Astronomy, Rutgers University, 136 Frelinghuysen Road, Piscataway, NJ 08854-8019}
\author{Kent~D.~Irwin}
\affiliation{Dept. of Physics, Stanford, CA 94305}
\affiliation{SLAC National Accelerator Laboratory, 2575 Sandhill Hill Road, Menlo Park, CA 94025}
\author{Brian~Koopman}
\affiliation{Department of Physics, Cornell University, Ithaca, NY 14853, USA}
\author{Arthur~Kosowsky}
\affiliation{Department of Physics and Astronomy, University of Pittsburgh, Pittsburgh, PA 15260 USA and
Pittsburgh Particle Physics, Astrophysics, and Cosmology Center, University of Pittsburgh, Pittsburgh PA 15260}
\author{Dale Li}
\affiliation{SLAC National Accelerator Laboratory, 2575 Sandhill Hill Road, Menlo Park, CA 94025}
\author{Thibaut~Louis}
\affiliation{Sub-Department of Astrophysics, University of Oxford, Keble Road, Oxford, UK OX1 3RH}
\author{Marius Lungu}
\affiliation{Department of Physics and Astronomy, University of
Pennsylvania, 209 South 33rd Street, Philadelphia, PA, USA 19104}
\author{Mathew Madhavacheril}
\affiliation{Physics and Astronomy Department, Stony Brook University, Stony Brook, NY USA 11794}
\author{Lo\"ic~Maurin}
\affiliation{Instituto de Astrof{\'{i}}sica, Pontific\'{i}a Universidad Cat\'{o}lica de Chile, Santiago, Chile}
\author{Jeffrey~John~McMahon}
\affiliation{Department of Physics, University of Michigan, Ann Arbor, USA 48103}
\author{Kavilan~Moodley}
\affiliation{Astrophysics and Cosmology Research Unit, School of Mathematics, Statistics and Computer Science, University of KwaZulu-Natal, Durban 4041, South Africa}
\author{Sigurd~Naess}
\affiliation{Sub-Department of Astrophysics, University of Oxford, Keble Road, Oxford, UK OX1 3RH}
\author{Federico~Nati}
\affiliation{Department of Physics and Astronomy, University of
Pennsylvania, 209 South 33rd Street, Philadelphia, PA, USA 19104}
\author{Laura Newburgh}
\affiliation{Dunlap Institute, University of Toronto, 50 St. George St., Toronto ON M5S3H4}
\author{Michael~D.~Niemack}
\affiliation{Department of Physics, Cornell University, Ithaca, NY 14853, USA}
\author{Lyman~A.~Page}
\affiliation{Joseph Henry Laboratories of Physics, Jadwin Hall,
Princeton University, Princeton, NJ, USA 08544}
\author{Christine G. Pappas}
\affiliation{Joseph Henry Laboratories of Physics, Jadwin Hall,
Princeton University, Princeton, NJ, USA 08544}
\author{Bruce~Partridge}
\affiliation{Department of Physics and Astronomy, Haverford College,
Haverford, PA, USA 19041}
\author{Benjamin L. Schmitt}
\affiliation{Department of Physics and Astronomy, University of
Pennsylvania, 209 South 33rd Street, Philadelphia, PA, USA 19104}
\author{Neelima Sehgal}
\affiliation{Physics and Astronomy Department, Stony Brook University, Stony Brook, NY USA 11794}
\author{Blake~D.~Sherwin}
\affiliation{Berkeley Center for Cosmological Physics, LBL and
Department of Physics, University of California, Berkeley, CA, USA 94720}
\affiliation{Miller Institute for Basic Research in Science, University of California, Berkeley, CA, 94720, USA}
\author{Jonathan~L.~Sievers}
\affiliation{Astrophysics and Cosmology Research Unit, School of Chemistry and Physics, University of KwaZulu-Natal, Durban 4041, South Africa}
\affiliation{National Institute for Theoretical Physics (NITheP), University of KwaZulu-Natal, Private Bag X54001, Durban 4000, South Africa}
\author{David N. Spergel}
\affiliation{Dept.~of Astrophysical Sciences, Peyton Hall, Princeton University, Princeton, NJ USA 08544}
\author{Suzanne~T.~Staggs}
\affiliation{Joseph Henry Laboratories of Physics, Jadwin Hall,
Princeton University, Princeton, NJ, USA 08544}
\author{Alexander van Engelen}
\affiliation{Canadian Institute for Theoretical Astrophysics, University of
Toronto, Toronto, ON, Canada M5S 3H8}
\author{Edward~J.~Wollack}
\affiliation{NASA/Goddard Space Flight Center, Greenbelt, MD, USA 20771}

%%%%%%%%%%%%%%%%%%%%%%%%%%%%%%%%%%%%%%%%%%%%%%%%%%%%%%%%%%%%%%%%%%%%%%%%%%%
\begin{abstract}
We use microwave temperature maps from two seasons of data from the Atacama Cosmology Telescope (ACTPol) at 146 GHz, together with the `Constant Mass' CMASS galaxy sample from the Baryon Oscillation Spectroscopic Survey to measure the kinematic Sunyaev-Zel'dovich (kSZ) effect over the redshift range $z=0.4-0.7$.
We use galaxy positions and the continuity equation to obtain a 
reconstruction of the line-of-sight velocity field. We stack the cosmic microwave background temperature at the location of each halo, weighted by the corresponding reconstructed velocity.
The resulting best fit kSZ model is preferred over the no-kSZ hypothesis at $3.3\sigma$ and $2.9\sigma$ for two independent velocity reconstruction methods, using $25,537$ galaxies over $660$ square degrees.
The effect of foregrounds that are uncorrelated with the galaxy velocities is expected to be well below our signal, and residual thermal Sunyaev-Zel'dovich contamination is controlled by masking the most massive clusters.
Finally, we discuss the systematics involved in converting our measurement of the kSZ amplitude into the mean free electron fraction of
the halos in our sample.

\end{abstract}
\pagebreak
\maketitle

%%%%%%%%%%%%%%%%%%%%%%%%%%%%%%%%%%%%%%%%%%%%%%%%%%%%%%%%%%%%%%%%%%%%%%%%%%%

\textit{Introduction.}
Measurements of the anisotropy in the Cosmic Microwave Background radiation (CMB), together with constraints from Big Bang Nucleosynthesis and the Lyman-$\alpha$ forest, tightly constrain the total baryon abundance of the Universe at $z \gtrsim 2$ \cite{2015arXiv150201589P, 2013ApJS..208...19H, 2014ApJ...781...31C}.
It is estimated that today, only about $10 \%$ of the baryons are found in stars or other neutral medium, while the majority of the rest are thought to occupy a warm, diffuse component \cite{2004ApJ...616..643F}. 
A large fraction of this component is believed to be in the form of the Warm-Hot Intergalactic Medium (WHIM), at typical temperatures of $10^5-10^7$ K, which is too cold and too diffuse to be visible with X-rays or through thermal Sunyaev-Zel'dovich (tSZ) observations, but at the same time too hot to collapse in dense cores and form stars \cite{2006ApJ...650..560C}. Due to the difficulty in observing the WHIM using current methods, the spatial distribution and abundance of baryons in the outskirts of galaxies and clusters is still poorly constrained, especially for group-sized objects or smaller.

The kinematic Sunyaev-Zel'dovich (kSZ) effect is the shift in CMB photon energy due to Thomson scattering off coherently moving electrons \cite{sun72, sun80}.  As we discuss below, the kSZ effect depends linearly on the local free electron density $n_e$, is independent of temperature $T_e$, and is therefore well-suited to probe the low density and low temperature outskirts of galaxies and clusters. This should be contrasted with the X-ray signal ($\propto n_e^2 \sqrt{T_e}$) and the tSZ signal ($\propto n_e T_e$), which receive their largest contributions from close to the cluster centers.

To lowest order, the kSZ effect is a Doppler shift, and therefore preserves the black body frequency spectrum of the CMB, simply shifting the brightness temperature.
In temperature units, the shift $\Delta T^{\rm kSZ}(\bm{\hat{n}})$ produced by the kSZ effect is sourced by the free electron \textit{momentum field}  $n_e \bm{v}_e$, and is given by \cite{sun72, 1986ApJ...306L..51O}
\be
\frac{\Delta T^{\rm kSZ}(\bm{\hat{n}})}{T_{\rm CMB}}  = -  \sigma_T \int \frac{d \chi}{1+z} e^{-\tau(\chi)} n_e(\chi\hat{\bm{n}},\chi) \ \frac{\bm{v}_e}{c} \cdot \bm{\hat{n}},
\label{eq:kSZdef}
\ee
where $\sigma_T$ is the Thomson scattering cross-section, $\chi(z)$ is the comoving distance to redshift $z$, $\tau$ is the optical depth to Thomson scattering, $n_e$ and $\bm{v}_e$ are the \textit{free} electron physical number density and peculiar velocity, and $\bm{\hat{n}}$ is the line-of-sight direction, defined to point away from the observer.
At late times, some fraction of the electrons in galaxies and clusters resides in the neutral medium (stars, stellar remnants, HI clouds, brown dwarfs etc.) and does not take part in the Thomson scattering that gives rise to the kSZ effect. 
We define $f_\text{free}$ as the fraction of free electrons compared to the expected cosmological abundance and note that the amplitude of the kSZ signal is directly proportional to it. The precise value of $f_{\rm free}$ is unknown and is expected to depend on redshift and mass; obtaining its value is one of the goals of precision kSZ measurements.
For an object with total mass (baryonic plus dark matter) $M_{\rm 200}$, we expect from Eq.~\eqref{eq:kSZdef}
$\Delta T^{\rm kSZ} \approx - 0.1 \mu {\rm K} \, f_{\rm free} \left( M_\text{200} / 10^{13} M_\odot \right)
\left( \bm{v}_e\cdot \bm{\hat{n}} / 300 \text{ km s}^{-1} \right)$, where we have taken the typical 1D RMS velocity at  $z\lesssim 0.5$ to be $300 \text{ km s}^{-1}$ and have defined $M_{200}$ to be the mass contained in a spherical volume with mean density 200 times the critical density at the halo redshift.

The kSZ signal is challenging to extract from the CMB, because a given halo can contribute a positive or negative signal with equal probability. Therefore a na\"{\i}ve stacking or cross-correlation analysis will lead to a large cancellation in the signal. To remedy this, a number of estimators have been proposed \cite{2009arXiv0903.2845H, 2011MNRAS.413..628S, 2014MNRAS.443.2311L, 1999ApJ...515L...1F, 2008PhRvD..77h3004B}. 
The first evidence for the kSZ signal was reported in \cite{Handetal2012} by using the pairwise velocity method, i.e. the fact that, on average, pairs of galaxies are moving toward rather than away from each other (see also \cite{2015arXiv150403339P}). Here we build upon the work of \cite{2013MNRAS.430.1617L, 2009arXiv0903.2845H, 2011MNRAS.413..628S, 2014MNRAS.443.2311L}, noting that if we have independent information on the peculiar velocity, we can weight halos by their velocities and avoid the cancellation (for a similar analysis, see \cite{2015arXiv150403339P, 2015arXiv150404011H}). 
Such estimates for the galaxy velocities can be obtained from the galaxy overdensity field by using the linearized continuity equation as described below. 

In this letter, we use CMB data from the Atacama Cosmology Telescope (ACTPol) \cite{2014JCAP...10..007N}, 
together with individual velocity estimates for the CMASS catalog of the Baryon Oscillation Spectroscopic Survey (BOSS) DR10 \cite{DR10} to provide evidence for the kSZ signal with signal to noise ratio $S/N=3.3$ and $2.9$, for the two independent reconstruction methods used.

%%%%%%%%%%%%%%%%%%%%%%%%%%%%%%%%%%%%%%%%%%%%%%%%%%%%%%%%%%%%%%%%%%%%%%%%%%%

\textit{Galaxy sample.}
CMASS galaxies have redshifts between $0.4$ and $0.7$ ($z_\text{median}=0.57$) \cite{2014MNRAS.441...24A}. A high fraction ($\sim 85\%$) of these galaxies resides at the center of galaxy groups or clusters \cite{2013MNRAS.432..743N} with mean total halo mass of $2\times 10^{13} M_\odot$ \cite{2013arXiv1311.1480M, 2014arXiv1407.1856M, 2015PhRvL.114o1302M}. 
The typical offset between the galaxy position and the halo center of mass is estimated to be $\lesssim 0.2'$ \cite{2012ApJ...757....2G}, 
much smaller than the $1.4'$ beam of the temperature map.
This makes CMASS galaxies excellent tracers of the center of their host halo. 

We use publicly available galaxy stellar mass estimates \cite{2013MNRAS.435.2764M}, obtained by fitting a stellar population synthesis model to the observed broadband spectral energy distribution (SED) of each CMASS galaxy.
These stellar masses range from $10^{11} M_\odot$ to $10^{12} M_\odot$, with a mean mass of $2\times 10^{11} M_\odot$.
The individual stellar mass estimates are converted to total masses for the host halos, following \cite{2014arXiv1401.7329K} (see also \cite{2013A&A...557A..52P}). 
Assuming cosmological baryon abundance (from Big Bang Nucleosynthesis \cite{2014ApJ...781...31C} or CMB \cite{2015arXiv150201589P}), we convert each halo mass into baryon mass. 
We assume that these baryons (hydrogen and helium with primordial abundance \cite{2013JCAP...11..017A, 2013A&A...558A..57I}) are fully ionized, which allows us to convert the baryon mass into the number of free electrons. This yields an estimate for the optical depth to Thomson scattering $\tau_i$ of each cluster $i$.
Note that these inferred optical depths are related by a factor of $1/f_{\rm free}$ to the true ones, 
since part of the electrons are in the neutral medium. This is taken into account consistently in the analysis.
A total of $25,537$ galaxies overlap with the ACTPol map and are included in the analysis.

%%%%%%%%%%%%%%%%%%%%%%%%%%%%%%%%%%%%%%%%%%%%%%%%%%%%%%%%%%%%%%%%%%%%%%%%%%%

\textit{Velocity reconstruction.}
A reconstructed velocity field can be inferred from the observed galaxy number overdensity $\delta_g$ by solving the linearized continuity equation in redshift space \cite{2012MNRAS.427.2132P}:
\be
\bm{\nabla}\cdot \bm{v}
+ f  \bm{\nabla}\cdot \left[ \left( \bm{v}\cdot \bm{\hat{n}} \right) \bm{\hat{n}} \right]
= -a H f \; \frac{\delta_g}{b}
\ee
where $f = d \ln \delta / d \ln a$ is the logarithmic linear growth rate.
Here we assumed that the galaxy overdensity $\delta_g$ is related to the total matter overdensity $\delta$ by a linear bias factor $b$, such that $\delta_g = b \, \delta$, with $b$ estimated from the auto-correlation of the galaxy catalog itself.

We use two different implementations of the velocity reconstruction:
the first one is used in the BOSS analysis for the purpose of Baryon Acoustic Oscillation peak reconstruction \cite{2012MNRAS.427.2132P, Mariana2015}.
The second one applies a Wiener filter to the galaxy number density field \cite{Smith2015}.
We refer to the two methods as VR1 and VR2 respectively.
Both implementations are tested on BOSS mock catalogs with realistic mask and selection function by comparing the `true' and reconstructed velocities. Using the PTHalos DR11 mock catalogs  \cite{2013MNRAS.428.1036M}, we find a correlation coefficient between true and reconstructed velocities of $r \simeq 0.65$ and $0.67$, 
and a multiplicative bias $\sigma_{v_\text{rec}} / \sigma_{v_\text{true}}$ of 0.64 and 0.69 for VR1 and VR2 respectively. The two methods are compared in detail in an upcoming paper \cite{Smith2015}.
Taking into account the properties of reconstructed velocities will be necessary for the interpretation of our measurement in terms of the physical properties of our sample, and this will be the subject of future work.

%%%%%%%%%%%%%%%%%%%%%%%%%%%%%%%%%%%%%%%%%%%%%%%%%%%%%%%%%%%%%%%%%%%%%%%%%%%
\textit{Microwave temperature maps.}
We use a map of the microwave intensity at $146$ GHz from ACTPol, a polarization sensitive receiver on the six meter Atacama Cosmology Telescope in Chile. 
Our map covers approximately $13^\circ$ in declination around the celestial equator, from right ascension $-10^\circ$ to $40^\circ$, and combines observations from ACT season 3 and 4 (2009 and 2010 data) \cite{2014JCAP...04..014D} and ACTPol season 1 and 2 (2013 and 2014 data) \cite{2014JCAP...10..007N}.
The effective beam full-width at half-maximum (FWHM) is $1.4'$, and the map noise level is approximately $14\mu$K$\cdot$arcmin, 
although it varies from $10\mu$K$\cdot$arcmin to $16\mu$K$\cdot$arcmin across the map.

An aperture photometry (AP) filter is applied at the position of each galaxy, and yields a noisy estimate $\delta T_i$ of the kSZ signal from the host halo. Applying the AP filter consists in averaging the value of the pixels within a disk of radius $\theta_{\rm disk}$, and subtracting the average of the pixels in an adjacent, equal area ring with external radius $\theta_{\rm ring} = \sqrt{2} \theta_{\rm disk}$. This estimate is dominated by primary CMB fluctuations (for aperture radii larger than $2'$) and map noise (for aperture radii smaller than $2'$), and is also affected by tSZ, galactic emission and other foregrounds. However, all these contaminants are uncorrelated with the cluster line-of-sight velocity and are expected to average out once weighted by the reconstructed velocities that have alternating sign. 

%%%%%%%%%%%%%%%%%%%%%%%%%%%%%%%%%%%%%%%%%%%%%%%%%%%%%%%%%%%%%%%%%%%%%%%%%%
\textit{Analysis.}
For each object in our sample, we combine the Thomson optical depth estimate $\tau_i$ and line-of-sight reconstructed velocity\footnote{Defined to be positive when pointing away from the observer.} $v_{{\rm rec}, i}$ with the AP filter output $\delta T_i$ at $146$~GHz. 
We define a number $\alpha$ as the best fit slope in the relation
\be
\frac{\delta T_i}{T_{\rm CMB}} = - \alpha \; \tau_i \frac{v_{{\rm rec}, i}}{c} \; .
\ee
Finding $\alpha$ consistent with zero means no detection of the kSZ effect, while finding $\alpha$ of order unity when the filter size is large enough to encompass the whole cluster corresponds to a number of free electrons consistent with the cosmological abundance. While $\alpha$ is directly proportional to the fraction of free electrons $f_{\rm free}$ within the filter, the proportionality coefficient is a non-trivial function of several variables (such as the filter size and shape, the baryon profile, the uncertainties in mass and velocity etc.). 
Accounting for these effects is required in order to constrain $f_\text{free}$ from our measurement, but is not necessary for the purpose of detection.

For each aperture size $\theta_{\rm disk}$, the best fit value of $\alpha$ is obtained by minimizing
\be
\sum_i \frac{\left( \delta T_i/T_{\rm CMB} + \alpha\; \tau_i v_{{\rm rec}, i} /c \right)^2}{\sigma_i^2} \; ,
\label{eq:chi2}
\ee
where the sum runs over all objects in our sample, and $\sigma_i^2$ is the variance of the filter output $\delta T_i$ caused by primary CMB fluctuations and noise\footnote{Here `noise' is taken to include not only detector noise, but all other effects that are uncorrelated with the signal, such as fluctuations in the atmosphere, and galactic and extragalactic foregrounds}. The value of $\sigma_i^2$ is determined by the size of the AP filter, as well as the level of map noise at the position of object $i$. The inverse-variance weighting $\propto 1/\sigma_i^2$ upweights the clusters that fall on less noisy parts of the CMB map. Minimizing Eq.~\eqref{eq:chi2} yields the best fit $\alpha$:
\be
\alpha =  - \frac{ \sum_i \left(\delta T_i /T_{\rm CMB} \right) \;  \left(\tau_i v_{{\rm rec}, i} /c\right)  / \sigma_i^2}
{ \sum_i \left(\tau_i v_{{\rm rec}, i}  /c\right)^2  / \sigma_i^2} .
\label{eq:alpha_best_fit}
\ee 

\begin{figure}[h]
\centering
\includegraphics[width=0.9\columnwidth]{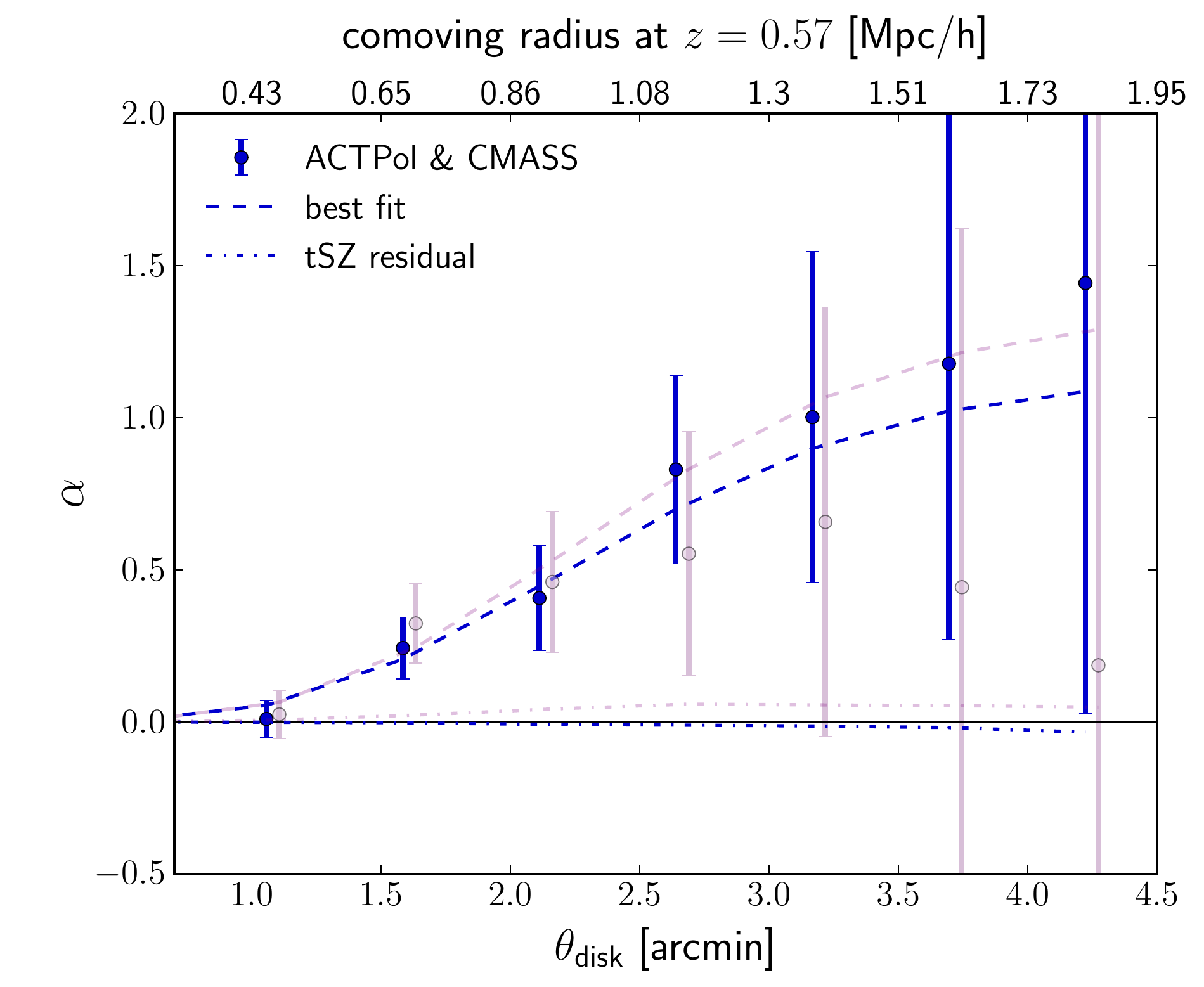}
\includegraphics[width=0.83\columnwidth]{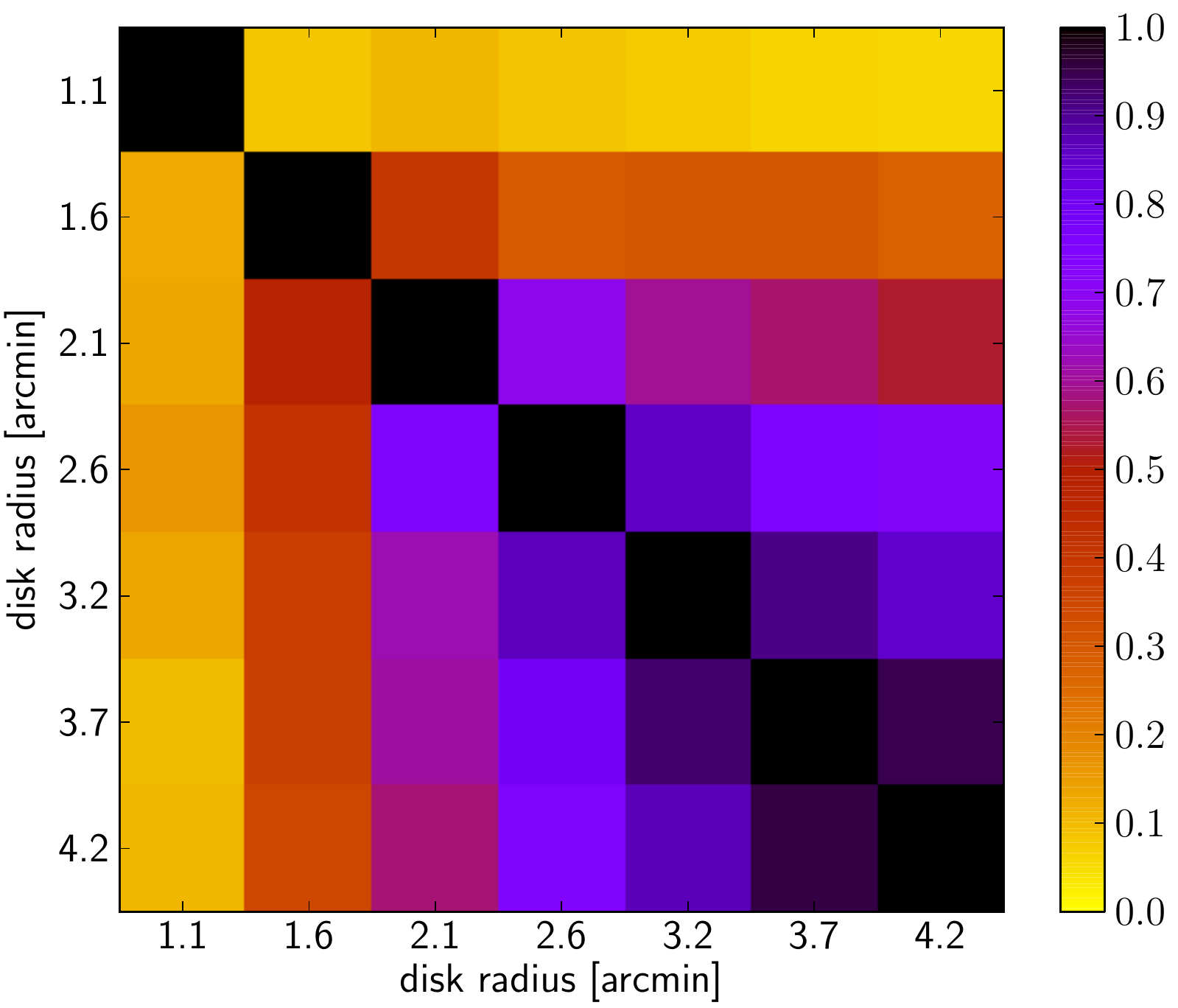}
\caption{
\textit{Top panel:}
Measured coefficient $\alpha$ (points with error bars) as a function of the angular radius $\theta_\text{disk}$ of the AP filter. The rate of change of $\alpha$ with $\theta_{\rm disk}$ is a proxy for the average baryon profile of our sample. 
The best fit curve (dashed line) is obtained by assuming a Gaussian projected profile with a scale of $1.5'$ (sum in quadrature of the beam and and the typical virial radius).
The tSZ residual (dot-dashed line) is negligible after masking the $1,126-2,881$ clusters more massive than $10^{14} M_\odot$.
The blue points and curves correspond to the velocity reconstruction method VR1, while the purple ones correspond to VR2. The kSZ signal is measured with $S/N = \sqrt{\Delta \chi^2} = 3.3$ for VR1 and $2.9$ for VR2.
\\
\textit{Bottom panel:}
Correlation coefficient matrix for the different aperture radii, for VR1 (above the diagonal) and VR2 (below the diagonal).
The data points are highly correlated, especially for the largest apertures. The signal to noise ratio is dominated by the three smallest apertures.
}
\label{fig:alpha}
\end{figure}

We repeat this analysis for various aperture radii. The best fit coefficient $\alpha$ is shown as a function of AP filter radius $\theta_{\rm disk}$ in Fig.~\ref{fig:alpha}. 
The various measurements of $\alpha$ for different $\theta_{\rm disk}$ are correlated since the data for a smaller $\theta_{\rm disk}$ is a subset of the data for a larger $\theta_{\rm disk}$.
In order to estimate the covariance matrix between the $\alpha$ for the various $\theta_\text{disk}$, we repeat the analysis above on 500 mock CMB maps, which include inhomogeneous noise due to the spatially varying depth of observation, as well as the observed power spectrum of foregrounds.
This method has the advantage of preserving the correlations in position and velocity for the BOSS objects, as well as the residual CMB correlations and the occasional overlap between the AP filters.

The CMASS halos have a typical angular size of $\theta_\text{vir} = 1.4'$, while the ACTPol beam is $\sigma_\text{beam} = 0.6'$ (corresponding to a FWHM of $1.4'$). Given the measurement uncertainties, it is reasonable to approximate the projected electron profile by a Gaussian of standard deviation $\sqrt{\theta_\text{vir}^2  + \sigma_\text{beam}^2} = 1.5'$. From this Gaussian profile, we predict the template for $\alpha$ as a function of $\theta_\text{disk}$, by applying the corresponding AP filters to the Gaussian profile. Intuitively, for small $\theta_\text{disk}$, the cluster kSZ signal contributes to the disk and the ring of the AP filter, which leads to a cancellation. For large $\theta_\text{disk}$, the cluster signal is entirely included in the disk of the AP filter, and the template goes to unity.
The dashed lines in Fig.~\ref{fig:alpha} correspond to this template, after fitting for an overall multiplicative amplitude.
We quantify the statistical significance (preference of the kSZ model over the ``no kSZ hypothesis'') as $S/N = \sqrt{\Delta \chi^2} = \sqrt{\chi^2_{\rm null} - \chi^2_{\rm bf}}$, where $\chi^2_{\rm null}$ and $\chi^2_{\rm bf}$ refer to the $\chi^2$ statistics applied to the null hypothesis and the best fit respectively. 
This signal to noise ratio is the inverse of the relative uncertainty on the best-fit amplitude.
We measure the kSZ signal with $S/N = 3.3 $ for VR1 and $2.9$ for VR2, with consistent amplitudes. 
We checked that numerical convergence errors on the covariance matrix affect the $\sqrt{\Delta \chi^2}$ value by less than $5\%$.

\begin{figure}[h]
\centering
\includegraphics[width=\columnwidth]{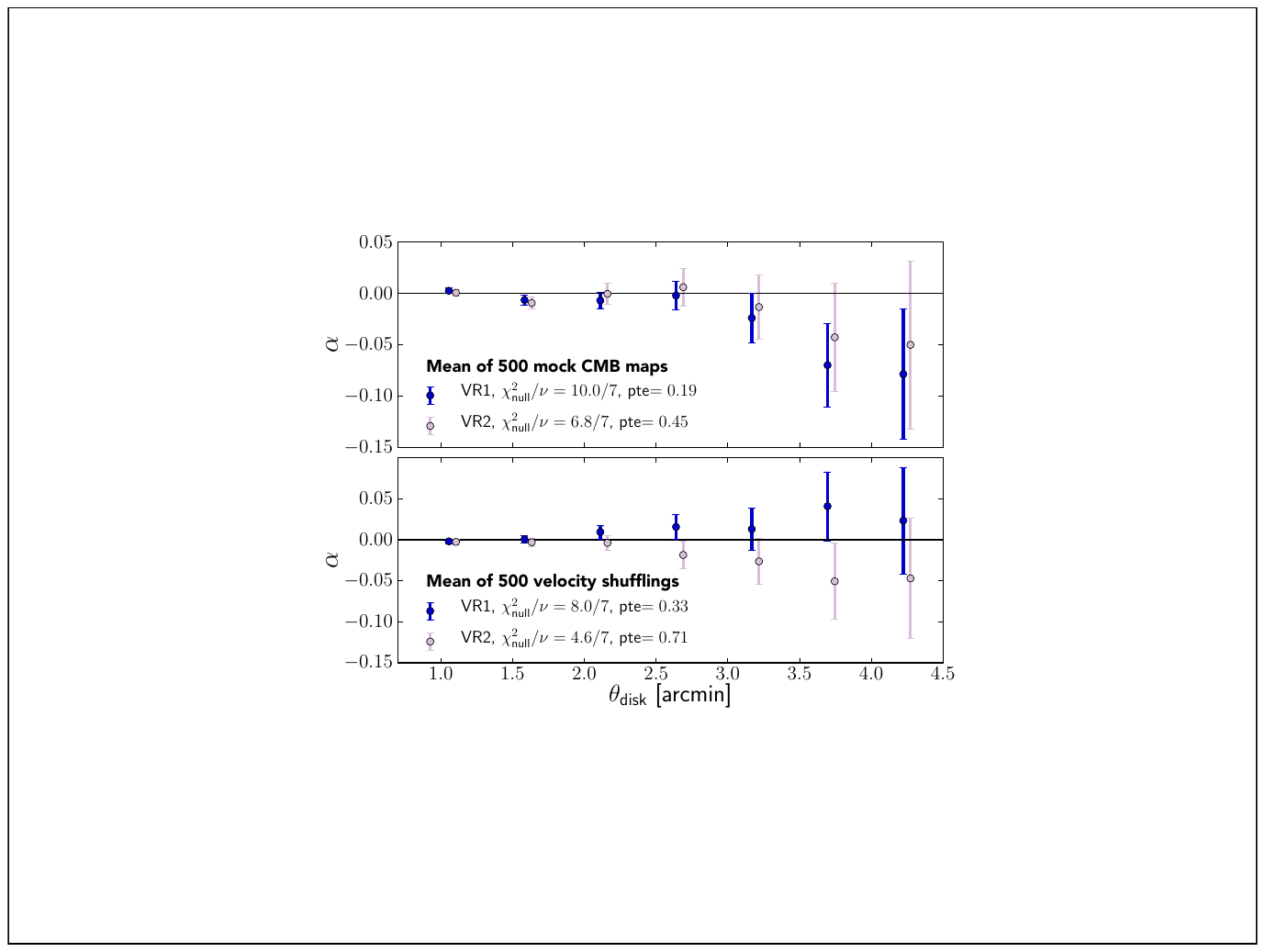}
\caption{
\textit{Top panel:} Mean of 500 null tests obtained by replacing the ACTPol map by a mock CMB map.\\
\textit{Bottom panel:} Mean of 500 null tests obtained by shuffling the reconstructed velocities among the CMASS objects.\\
Both panels show that the kSZ signal is only detected when analyzing the true ACTPol map and when assigning the correct velocity to each cluster.
}
\label{fig:null_tests}
\end{figure}

%%%%%%%%%%%%%%%%%%%%%%%%%%%%%%%%%%%%%%%%%%%%%%%%%%%%%%%%%%%%%%%%%%%%%%%%%%%

\textit{Null tests and systematics.}
A number of null tests are performed, as shown in Fig.~\ref{fig:null_tests}.
The procedure described to estimate the covariance matrix provides a first null test.
It shows that the kSZ signal is only detected when analyzing the true temperature map, which means that the signal is not due to unexpected features of the galaxy catalog.
We further confirm that the kSZ signal is only detected when the correct velocity is attributed to each object,
by shuffling the velocities $v_{{\rm rec},i}$ among the clusters in our sample.
In all cases, the kSZ signal disappears and the result becomes consistent with the null hypothesis.

The kSZ signal is proportional to the halo mass, while the tSZ signal is proportional to a higher power of mass (about $M_\text{halo}^{5/3}$). Thus the tSZ signal is typically larger than the kSZ signal for massive clusters\footnote{At 146 GHz, we find that for an object with line-of-sight velocity equal to the 1D rms, the tSZ signal becomes larger than the kSZ signal for $M_{200} > 2 \times 10^{13} M_{\odot}$.}. The contamination of $\alpha$ is mitigated by the fact that the tSZ signal is uncorrelated with the line-of-sight velocity and is weighted by alternate signs (see Eq.~\eqref{eq:alpha_best_fit}). However for very rare objects this cancellation may be incomplete and can be a significant source of contamination.
We estimate the size of the tSZ contamination to the value of $\alpha$ by replacing the measured cluster temperatures $\delta T_i$ by estimates for their tSZ signal \cite{2013A&A...557A..52P, 2015ApJ...808..151G} based on their stellar masses. We find the tSZ contamination to be important when including clusters with total mass greater than a few $\times 10^{14} M_\odot$. Indeed, these objects are rare enough that the cancellation in the numerator of Eq.~\eqref{eq:alpha_best_fit} is incomplete. Masking objects with  $M_{200} > 10^{14} M_\odot$, together with a $1'$ region around them, is sufficient to limit the tSZ contamination to less than 10\% of the statistical uncertainty on $\alpha$. This removes 1,126 objects (for the smallest AP size) to 2,881 objects (for the largest AP size) from the analysis. 
We assess the amplitude of extragalactic thermal dust contamination from these objects by stacking the CMB map (with uniform weight) at the object positions and conclude that dust emission at 146 GHz is either negligible for our purposes (for low mass halos), or subdominant to tSZ (for high mass objects). Therefore dust is not expected  to be a significant contaminant after masking the most massive clusters.\footnote{Galactic dust and the bulk of the Cosmic Infrared Background (CIB) emission are uncorrelated with the CMASS galaxy positions and therefore are only an additional source of noise, which is included in the covariance matrix.}

Our analysis pipeline is tested on realistic mock kSZ realizations: a kSZ template is obtained by populating BOSS mock catalogs (PTHalos DR11 \cite{2013MNRAS.428.1036M}) of galaxy positions and velocities with Gaussian cluster profiles and then added to the CMB map, which provides the correct noise level. These mock maps are then analyzed the same way as the real data, by using both the `real' and `reconstructed' velocities, obtaining consistent results. The loss in signal-to-noise when using the reconstructed rather than the real velocities is equal to the correlation coefficient $r$ as expected.
We estimate the effect of cluster miscentering by adding an offset of $0.2'$ (which is roughly the expected rms miscentering \cite{2012ApJ...757....2G}) to the cluster centers in the mocks. This leads to less than 3\% change in $\alpha$.

%%%%%%%%%%%%%%%%%%%%%%%%%%%%%%%%%%%%%%%%%%%%%%%%%%%%%%%%%%%%%%%%%%%%%%%%%%%
\textit{Interpretation.}
We have presented evidence for the kSZ signal with overall $S/N \simeq 3$.
We defined a coefficient $\alpha$ as the best fit proportionality constant between the AP filter output and the expected kSZ signal.
This number $\alpha$ can only be interpreted as the free electron fraction $f_\text{free}$ if all of the electrons associated with each cluster are within the filter aperture, if there is no effect from galaxy overlap, if all the galaxies in our sample are central galaxies and if both the velocities and masses are known exactly.
This is clearly not the case here, so the physical interpretation of $\alpha$ is not straightforward. We now briefly discuss these effects, which determine the relationship between $\alpha$ and $ f_{\rm free}$, and defer a careful and in-depth analysis of these effects to upcoming work. 

If the kSZ emission from the object does not entirely fall within the inner disk of the AP filter, part of the signal will be subtracted off, reducing the observed value of $\alpha$. This is clearly visible in Figure 1, for small $\theta_{\rm disk}$: the size of the disk for $\theta_{\rm disk} \ll 2'$ is smaller than the extent of the emission and the signal is cancelled by the surrounding ring. 
For large apertures $\theta_{\rm disk}$, we expect this cancellation to disappear and $\alpha$ to asymptote to $f_{\rm free}$.

The gas spatial profile would then determine the rate of increase of $\alpha$ from 0 to $f_\text{free}$. In fact, Figure \ref{fig:alpha} can be though of a proxy for the average baryon profile of our sample. However, the noise from primary CMB fluctuations also increases with $\theta_{\rm disk}$, making it difficult to disentangle the free electron fraction $f_{\rm free}$ from the spatial size of the cluster. 

The reconstructed velocities are biased low and are not 100\% correlated with the true velocities. Therefore, $\alpha$ differs from $f_{\rm free}$ by an additional factor of $r \sigma_{v_\text{true}} / \sigma_{v_\text{rec}}$, as can be inferred from Eq.~\eqref{eq:alpha_best_fit}. This factor can be measured from mock catalogs where true and reconstructed velocities can be compared.

We use an average stellar mass to halo mass relation. 
The typical intrinsic scatter in this relation \cite{2013A&A...557A..52P, 2014arXiv1401.7329K}, as well as potential errors on the stellar mass determination, can lead to a bias in $\alpha$ of up to $\sim$ 40\%. 

The presence of extra free electrons with correlated velocities (unbound or associated with a different cluster) within a single aperture is expected to bias $\alpha$ high. This effect can be interpreted as a 2-halo term in the kSZ correlation function, where the presence of additional mass correlated with the galaxies used for stacking contributes a signal at large enough separations.

\textit{Outlook.}
As the overlap between large-scale structure datasets and high sensitivity CMB maps increases,
the significance of kSZ detections will see a rapid improvement.
Future surveys such as Advanced ACTPol \cite{2015arXiv151002809H} and SPT-3G \cite{2014SPIE.9153E..1PB} should enable a few percent-level precision kSZ measurement.

Combined with a better understanding of the relationship between the observed signal and the underlying physical properties of the sample, these high-significance detections will enable a precise measurement of the free electron fraction and the baryon profile of the low-density regions in the outskirts of galaxies and clusters, which are believed to host the majority of the gas.

These measurements can be performed as a function of mass and redshift, and combined with tSZ and X-ray observations of the same objects to independently measure density and temperature profiles. These measurements will shed new light on galaxy evolution and feedback processes within clusters, which can be used to improve the cosmological constraints from cluster counts \cite{2013JCAP...07..008H, 2015arXiv150201597P} and our understanding of the matter power spectrum on small scales \cite{2008ApJ...672...19R, 2015arXiv150102055O}.

Once the astrophysical quantities are well-characterized, the kSZ signal itself can also be used for a number of cosmological applications, such as constraining bulk flows \cite{2012ApJ...758....4M, 2014A&A...561A..97P}, probing neutrino physics \cite{2014arXiv1412.0592M} and testing general relativity \cite{2009PhRvD..80f2003K, 2014arXiv1408.6248M}.

%%%%%%%%%%%%%%%%%%%%%%%%%%%%%%%%%%%%%%%%%%%%%%%%%%%%%%%%%%%%%%%%%%%%%%%%%%%
\textit{Acknowledgements.} 
We thank Marcelo Alvarez, Neal Dalal, Tommaso Giannantonio, Oliver Hahn, Andrey Kravtsov, Guilhem Lavaux, Hironao Miyatake, Hyunbae Park, Hiranya Peiris, Ue-Li Pen, Bjoern Soergel, Naonori Sugiyama and Simon White for very useful discussion.

This work was supported by the U.S. National Science Foundation through awards
AST-0408698 and AST-0965625 for the ACT project, as well as awards PHY-0855887
and PHY-1214379. Funding was also provided by Princeton University, the
University of Pennsylvania, Cornell University and a Canada Foundation for Innovation (CFI) award
to UBC. ACT operates in the Parque Astron\'omico Atacama in northern Chile
under the auspices of the Comisi\'on Nacional de Investigaci\'on Cient\'ifica y
Tecnol\'ogica de Chile (CONICYT). Computations were performed on the GPC
supercomputer at the SciNet HPC Consortium. SciNet is funded by the CFI under
the auspices of Compute Canada, the Government of Ontario, the Ontario Research
Fund -- Research Excellence; and the University of Toronto. 
Colleagues at RadioSky provide logistical support and keep operations in
Chile running smoothly. We also thank the Mishrahi Fund and the
Wilkinson Fund for their generous support of the project.

ES, SF and DNS are supported by NSF grant AST1311756 and NASA grant NNX12AG72G.
Research at Perimeter Institute is supported by the Government of Canada through Industry Canada and by the Province of Ontario through the Ministry of Research and Innovation. KMS was supported by an NSERC Discovery Grant.
SH is supported in part by DOE-ASC DOE-DESC0011114, NASA 12-EUCLID11-0004, NSF AST1517593 and NSF AST1412966.
NB acknowledges support from the Lyman Spitzer Fellowship.
MN and FDB acknowledge support from NSF grants AST-1454881 and AST- 1517049.

The development of multichroic detectors and lenses was supported by
NASA grants NNX13AE56G and NNX14AB58G. CM acknowledges support from NASA
grant NNX12AM32H. Funding from ERC grant 259505 supports SN, JD, EC, and
TL. HT is supported by grants NASA ATP NNX14AB57G, DOE DE-SC0011114, and
NSF AST- 1312991. BS and BK are funded by NASA Space Technology
Research Fellowships. R.D received funding from the CONICYT grants
QUIMAL-120001 and FONDECYT-1141113.

This research used resources of the National Energy Research Scientific Computing Center, a DOE Office of Science User Facility supported by the Office of Science of the U.S. Department of Energy under Contract No. DE-AC02-05CH11231.

Funding for SDSS-III has been provided by the Alfred P. Sloan Foundation, the Participating Institutions, the National Science Foundation, and the U.S. Department of Energy Office of Science. The SDSS-III web site is http://www.sdss3.org/.
SDSS-III is managed by the Astrophysical Research Consortium for the Participating Institutions of the SDSS-III Collaboration including the University of Arizona, the Brazilian Participation Group, Brookhaven National Laboratory, Carnegie Mellon University, University of Florida, the French Participation Group, the German Participation Group, Harvard University, the Instituto de Astrofisica de Canarias, the Michigan State/Notre Dame/JINA Participation Group, Johns Hopkins University, Lawrence Berkeley National Laboratory, Max Planck Institute for Astrophysics, Max Planck Institute for Extraterrestrial Physics, New Mexico State University, New York University, Ohio State University, Pennsylvania State University, University of Portsmouth, Princeton University, the Spanish Participation Group, University of Tokyo, University of Utah, Vanderbilt University, University of Virginia, University of Washington, and Yale University.

%%%%%%%%%%%%%%%%%%%%%%%%%%%%%%%%%%%%%%%%%%%%%%%%%%%%%%%%%%%%%%%%%%%%%%%%%%%

\end{document}